\documentclass[12pt,fleqn]{article}
\usepackage{graphics,graphicx,cite}
\usepackage{amssymb,amsfonts}

\def\onecol{\onecolumn \mathindent 2em}

\def\noi{\noindent}

\newcommand{\sect}[1]{Sec.\,#1}

\def\nhh{\hspace*{-0.3em}}
\def\cm{\hspace*{1cm}}

\def\Jl#1#2{{\it #1\/} {\bf #2},\ }

\def\ApJ#1 {\Jl{Astroph. J.}{#1}}
\def\CQG#1 {\Jl{Class. Quantum Grav.}{#1}}
\def\DAN#1 {\Jl{Dokl. AN SSSR}{#1}}
\def\GC#1 {\Jl{Grav. \& Cosmol.}{#1}}
\def\GRG#1 {\Jl{Gen. Rel. Grav.}{#1}}
\def\JETF#1 {\Jl{Zh. Eksp. Teor. Fiz.}{#1}}
\def\JETP#1 {\Jl{Sov. Phys. JETP}{#1}}
\def\JHEP#1 {\Jl{JHEP}{#1}}
\def\JMP#1 {\Jl{J. Math. Phys.}{#1}}
\def\NPB#1 {\Jl{Nucl. Phys.}{B\ #1}}
\def\NP#1 {\Jl{Nucl. Phys.}{#1}}
\def\PLA#1 {\Jl{Phys. Lett.}{#1A}}
\def\PLB#1 {\Jl{Phys. Lett.}{#1B}}
\def\PRD#1 {\Jl{Phys. Rev.}{D\ #1}}
\def\PRL#1 {\Jl{Phys. Rev. Lett.}{#1}}

\def\lal{&& {}\nhh}
\def\eq{Eq.\,}
\def\eqs{Eqs.\,}
\def\beq{\begin{equation}}
\def\eeq{\end{equation}}
\def\bear{\begin{eqnarray}}
\def\bearr{\bear \lal}
\def\ear{\end{eqnarray}}

\def\nn{\nonumber\\ {}}

\def\eql{&\! = &\!}
\def\sequ#1{\setcounter{equation}{#1}}

\def\dst{\displaystyle}
\def\tst{\textstyle}
\def\fracd#1#2{{\dst\frac{#1}{#2}}}
\def\fract#1#2{{\tst\frac{#1}{#2}}}
\def\Half{{\fracd{1}{2}}}
\def\half{{\fract{1}{2}}}

\def\e{{\,\rm e}}
\def\d{\partial}

\def\sign{\mathop{\rm sign}\nolimits}

\def\const{{\rm const}}
\def\eps{\varepsilon}

\newcommand{\vars}[1]{\left\{\begin{array}{ll}#1\end{array}\right.}

\def\M{{\mathbb M}}
\def\N{{\mathbb N}}
\def\R{{\mathbb R}}
\def\S{{\mathbb S}}

\def\ME {\mbox{$\M_{\rm E}$}}
\def\MJ {\mbox{$\M_{\rm J}$}}

\def\cF{{\cal F}}

\def\og{\overline{g}}
\def\cR{{\cal R}}
\def\ocR{\overline{\cal R}}

\def\mn{_{\mu\nu}}
\def\MN{^{\mu\nu}}
\def\mN{_\mu^\nu}

\def\TH{\mbox{$T_{\rm H}$}}
\def\Str{\mbox{$\S_{\rm trans}$}}

\def\wh{wormhole}
\def\whs{wormholes}
\def\bh{black hole}
\def\bhs{black holes}

\def\ssph{static, spherically symmetric}
\def\asflat{asymptotically flat}

\def\Kr{Kretsch\-mann}

\def\BD{Brans-Dicke}

\begin{document}
\onecol

\title{\bf Cold black holes and conformal continuations}

{\small
\author{
   K.A. Bronnikov\footnote{E-mail: kb20@yandex.ru}\\
   Centre for Gravitation and Fundamental Metrology, VNIIMS,\\
    46 Ozyornaya St., Moscow 119461, Russia;\\
     Institute of Gravitation and Cosmology, PFUR,\\
     6 Miklukho-Maklaya St., Moscow 117198, Russia\\
\\
 M.S. Chernakova \\
     Institute of Gravitation and Cosmology, PFUR,\\
     6 Miklukho-Maklaya St., Moscow 117198, Russia\\
\\
J.C. Fabris\footnote{E-mail: fabris@cce.ufes.br}\\
Departamento de F\'{\i}sica, Universidade Federal do Esp\'{\i}rito Santo,\\
      Vit\'oria, 29060-900, Esp\'{\i}rito Santo, Brazil\\
\\
       N. Pinto-Neto\footnote{E-mail: nelsonpn@cbpf.br}
    \ and M.E. Rodrigues\footnote{E-mail: esialg@gmail.com}\\
    ICRA-CBPF, Rua Xavier Sigaud, 150, Urca, CEP22290-180,
        Rio de Janeiro, Brazil} \date{}
}
\maketitle
\begin{abstract}
  We study Einstein gravity minimally coupled to a scalar field in a static,
  spherically symmetric space-time in four dimensions. Black hole solutions
  are shown to exist for a phantom scalar field whose kinetic energy
  is negative. These ``scalar black holes'' have an infinite horizon area
  and zero temperature \TH\ and are termed ``cold black holes'' (CBHs).
  The relevant explicit solutions are well-known in the massless case (the
  so-called anti-Fisher solution), and we have found a particular example of
  a CBH with a nonzero potential $V(\phi)$. All CBHs with $V(\phi)
  \not\equiv 0$ are shown to behave near the horizon quite similarly to
  those with a massless field. The above solutions can be converted by a
  conformal transformation to Jordan frames of a general class of
  scalar-tensor theories of gravity, but CBH horizons in one frame are in
  many cases converted to singularities in the other, which gives rise to a
  new type of conformal continuation.


\end{abstract}

\newpage

\section{Introduction}

  The conventional idea of a black hole (BH) implies a space-time singularity
  hidden beyond an event horizon \cite{misner}, a hypersurface which
  separates an external region, containing spatial infinity, from an
  internal region, invisible to an external observer. The well-known BH
  solutions of general relativity (Schwarzschild, Reissner-Nordstr\"om,
  Kerr, Kerr-Newman), have been generalized in many contexts, such as the
  presence of scalar fields of various nature, non-linear gravity theories,
  scalar-tensor theories etc. (see, e.g., \cite{lemos} and references
  therein). Their different properties rise the question of an extension of
  the \bh\ notion itself. An example of ``exotic'' black holes are the
  so-called ``cold black holes'', obtained in scalar-tensor theories
  (STT) in general and in the Brans-Dicke theory in particular
  \cite{lousto,k1,k2}.

  The static, spherically symmetric solutions of the Brans-Dicke theory
  reveal a large class of objects with black hole properties. Not all of
  them exhibit a singularity beyond a horizon. However, the horizon surface
  has in all such cases an infinite area. Moreover, all such horizons have
  zero surface gravity and hence zero Hawking temperature.  It is for this
  reason that they have been named ``cold black holes'' (CBHs).

  As other STT, the Brans-Dicke theory can be re-defined using a conformal
  mapping to the so-called Einstein frame, or picture, in which the
  nonminimal coupling between the scalar field and the curvature, which is
  an essential feature of an STT, is ruled out, resulting in Einstein
  gravity with a minimally coupled scalar field. The scalar field energy is
  positive (its kinetic term has its usual sign) if the \BD\ coupling
  constant $\omega > - 3/2$, and is negative if $\omega < - 3/2$. In the
  latter case, the kinetic term has a ``wrong'' sign, and the theory is
  called anomalous, or phantom. Such kind of theories have recently become
  quite fashionable for both theoretical and observational reasons. The
  theoretical reasons are connected with the ghost condensation and
  tachyonic fields that result from string theories \cite{piazza,bagla}.
  From the observational viewpoint, recent analysis of the type Ia supernova
  and CMB data indicates that perhaps the best fit is given by {\it phantom
  fields} \cite{caldwell,hannestad,alam,allen}, of which a scalar field with
  the ``wrong'' sign of the kinetic term is the simplest example.

  In the present work, we show that, as opposed to what has been believed
  \cite{xanthopoulos}, this Einstein-scalar field system, with a massless
  scalar field minimally coupled to gravity, admits black hole solutions,
  though this happens for a phantom scalar field only.  These ``scalar''
  black holes also have an infinite horizon area and zero temperature.
  Such solutions can be interpreted as Einstein-frame solutions of any STT
  and transformed to the Jordan frame by the inverse conformal mapping.

  However, an interesting aspect of this procedure is the non-existence of a
  one-to-one correspondence between black holes in the Einstein and Jordan
  pictures, as we shall see using the \BD\ theory as an example. The reason
  is that, generically, the above conformal mapping (direct or inverse)
  converts a CBH horizon to a singularity. Such a situation is an example of
  a {\it conformal continuation\/} (CC). This phenomenon was treated in some
  detail in the framework of STT and $f(R)$ theories of gravity in
  Refs.\,\cite{vac3,vac4,hog1,hog2}. The point is that a conformal mapping
  between two manifolds $\M_1$ and $\M_2$ comprises a one-to-one
  correspondence between the respective points and preserves the causal
  structure only if the conformal factor is everywhere smooth and finite.
  If, however, the conformal factor somewhere becomes infinite or zero, the
  mapping, in general, links only a portion of $\M_1$ to a portion of
  $\M_2$. In some cases, a singularity in $\M_1$ may be mapped into a regular
  surface in $\M_2$, then $\M_2$ continues beyond this surface and may have
  more complex global properties as compared to $\M_1$. The new region may,
  in particular, contain a singularity, a horizon or another spatial
  infinity.

  From a more general point of view, the possible existence of CCs may mean
  that the observed Universe is only a region of a real, greater Universe
  which should be described in another, more fundamental conformal frame
  than the one related to our measurement instruments. Detailed discussions
  of the physical meaning and role of different conformal frames in the
  description of the Universe may be found in Refs.\,\cite{fara,bm-erice}.

  We shall find in this study that the appearance of CCs in the context of
  \ssph\ solutions of STT is closely related to the occurrence of a peculiar
  type of space-time singularities, where all curvature invariants remain
  finite but the analyticity of the metric is lost, which means that the
  manifold terminates. It is this kind of singularities that are in many
  cases removed by conformal mappings, leading to a CC.

  We here restrict ourselves to the search and discussion of CBH solutions
  in the Einstein-scalar field system. A more general analysis of the same
  system with nonzero potentials $V(\phi)$ (but without discussing
  infinite-area horizons) has been performed in Ref.\,\cite{pha1}, where it
  was shown that phantom scalar fields with appropriate potentials can form
  as many as sixteen types of regular \ssph\ self-gravitating
  configurations, including regular black holes with nonzero temperature.

  The paper is organized as follows. In the next section, we reproduce the
  basic equations of a general STT for \ssph\ metrics. \sect 3 describes the
  basic properties of the static, spherically symmetric solutions to the
  Einstein-massless scalar field system, which simultaneously represent the
  Einstein-frame solutions to a general class of STT with zero scalar field
  potential. We single out a particular discrete family of solutions
  corresponding to CBHs. In \sect 4, we compare the STT solutions in the
  Einstein and Jordan conformal frames, using as an example the \BD\
  theory, and pay special attention to CBH solutions in both frames. We
  discuss a new type of conformal continuations that appears in this
  context and make some remarks on the thermodynamical properties of CBHs,
  e.g., concerning the conformal invariance of the Hawking temperature.
  \sect 5 is devoted to scalar-vacuum configurations with nonzero potentials
  $V(\phi)$. We show that the nature of CBH horizons is basically the same
  for both zero and nonzero $V(\phi)$ and present a specific example of a
  CBH with $V(\phi)\not\equiv 0$. In \sect 5, we formulate our conclusions,
  and, finally, in the Appendix we present some general relations for \ssph\
  metrics and show, in particular, that horizons with an infinite area
  always possess zero Hawking temperature.

\section{Scalar-tensor theory: basic equations}

    In the general (Bergmann-Wagoner-Nordtvedt) 4-dimensional
    STT, the action in the pseudo-Riemannian manifold $\MJ[g]$ has the form
\bearr
     S_{\rm STT} = \int d^4 x \sqrt{g}                       \label{act-J}
           [f(\varphi) \cR
           + h(\varphi) (\d\varphi)^2 -2U(\varphi) + L_m],
\ear
    where $g\mn$ is the metric, $\cR=\cR[g]$ is the scalar curvature,
    $g=|\det g\mn|$, $f$, $h$ and $U$ are functions of the real scalar field
    $\varphi$, $(\d\varphi)^2 = g\MN\d_\mu\varphi\d_\nu\varphi$, and $L_m$
    is the matter Lagrangian. The manifold $\MJ[g]$ with the metric $g\mn$
    comprises the so-called Jordan conformal frame.

    The standard transition to the Einstein frame $\ME[\og]$ \cite{wagon},
\bear
    g\mn \eql |f(\varphi)|^{-1} \og\mn,         \label{g-wag}
\\
    \frac{d\phi}{d\varphi} \eql \frac{\sqrt{|l(\varphi)|}}{f(\varphi)},
\cm                                                     \label{phi-wag}
    l(\varphi) = fh + \frac{3}{2}\biggl(\frac{df}{d\varphi}\biggr)^2,
\ear
    removes the nonminimal scalar-tensor coupling express\-ed in a
    $\varphi$-dependent coefficient before $\cR$. Putting $L_m=0$ (vacuum),
    one can write the action (\ref{act-J}) in terms of the new metric
    $\og\mn$ and the new scalar field $\phi$ as follows (up to a boundary
    term):
\bearr
    S_{\rm E} = \int  d^4 x \sqrt{\og}                   \label{act-E}
    \Bigl\{(\sign f) \bigl[\ocR
                + (\sign l) (\d\phi)^2\bigr] - 2V(\phi)\Bigr\},
\ear
    where the determinant $\og$, the scalar curvature $\ocR$ and $(\d\phi)^2$
    are calculated using $\og\mn$ and
\beq
       V(\phi) = |f|^{-2}\,U(\varphi).               \label{V-E}
\eeq

    Note that $\eps := \sign l = -1$ corresponds to the so-called anomalous
    STT, with a wrong sign of scalar field kinetic energy in the Einstein
    frame, while $\sign f = -1$ means that the effective gravitational
    constant in the Jordan frame (which can be defined as $1/f$ up to a
    constant factor) is negative. So the normal choice of signs is $\sign l
    = \sign f =1$. Nevertheless, theories admitting $\eps = -1$ and/or
    $f < 0$ possess many features of interest, worth studying.
    With $f > 0$, the action (\ref{act-E}) describes Einstein gravity
    minimally coupled to a self-interacting scalar field, which is called
    normal for $\eps = +1$ and phantom for $\eps =-1$.
    The field equations are
\bear
       R\mn \eql  \eps \phi_{;\mu}\phi_{;\nu} - g_{\mu\nu}V(\phi) ,
\nn
       \nabla^\mu \nabla_\mu \phi \eql  - \eps dV/d\phi.
\ear

    Consider the general static, spherically symmetric metric in \ME
\beq                                                        \label{ds_u}
        ds^2 = \e^{2\gamma}dt^2 - \e^{2\alpha}du^2 - \e^{2\beta}d\Omega^2,
\eeq
    where $\gamma = \gamma(u)$, $\alpha = \alpha(u)$ and $\beta = \beta(u)$,
    $u$ is an arbitrary radial coordinate. Let us assume $f > 0$ but admit
    $\eps = \pm 1$. The field equations are
\bear                                   \label{e1}
    \gamma'' + \gamma'(\gamma' - \alpha' + 2\beta')
                    \eql  - \e^{2\alpha}V(\phi) ,
\\                                  \label{e2}
     \gamma'' + 2\beta'' + \gamma'^2 + 2\beta'^2 - \alpha'(\gamma' +
    2\beta') \eql  - \eps {\phi'^2} - \e^{2\alpha}V(\phi) ,
\\                                  \label{e3}
    \beta'' + \beta'(\gamma' - \alpha'+ 2\beta') - \e^{2\alpha - 2\beta}
                    \eql  - \e^{2\alpha}V(\phi) ,
\\                                      \label{e4}
    \bigr(\e^{\gamma - \alpha + 2\beta}\phi'\bigl)'
        \eql  \eps \e^{\gamma + 2\beta + \alpha} dV/d\phi ,
\ear
    where the prime denotes $d/du$. \eqs (\ref{e1})--(\ref{e4}) can be
    simplified by making a special choice of the radial coordinate $u$.
    Thus, choosing the quasiglobal coordinate $u = \rho$ defined by the
    condition $\alpha + \gamma =0$ (it is particularly convenient for
    considering metrics with Killing horizons) and denoting
\beq
    \e^{2\gamma} = \e^{-2\alpha} \equiv A(\rho) , \cm
                          \e^\beta = r(\rho),
\eeq
    it is straightforward to bring \eqs (\ref{e1})--(\ref{e4}) to the form
\bear                                                         \label{ne1}
    (A\,r^2\,\phi')' \eql  \eps r^2\,V_\phi ,
\\                                                            \label{ne2}
    (A'\,r^2)'\eql  - 2V\,r^2 ,
\\                                                            \label{ne3}
    2r''/r \eql  - \eps\,\phi'^2 ,
\\                                                            \label{ne4}
    (r^2)''\,A - A''\,r^2 \eql  2.
\ear
    \eq (\ref{ne4}) is once integrated giving
\beq                                                          \label{ne4i}
    (A/r^2)' = 2 (\rho_0 - \rho)/r^4, \cm \rho_0 = \const.
\eeq

    Solutions to these equations with $V(\phi) \equiv 0$ are well known
    \cite{fisher,Afish,br73}. For $V(\phi) \not\equiv 0$, a general analysis
    has been performed in \cite{pha1}, establishing the existence of regular
    configurations including BHs which can contain one or two horizons with
    finite area. Complete analytical solutions can only be found for special
    choices of $V(\phi)$.

    Given a solution to \eqs (\ref{e1})--(\ref{e4}) or
    (\ref{ne1})--(\ref{ne4}), the corresponding solution in the original,
    Jordan picture is easily obtained using the inverse transformation
    (\ref{g-wag}).

\section {Fisher and anti-Fisher solutions}   

    Let us reproduce the well-known solutions to \eqs (\ref{e1})--(\ref{e4})
    for zero potential, $V \equiv 0$. In case $\eps = +1$ the solution was
    found by I.Z. Fisher in 1947 \cite{fisher} and afterwards repeatedly
    re-discovered. For $\eps = -1$ the corresponding solution was first
    obtained, to our knowledge, by Bergmann and Leipnik \cite{Afish}. However,
    these authors used the curvature coordinates [i.e., the condition
    $u\equiv r$ in terms of the metric (\ref{ds_u})] which are not well
    suited for the problem, and this was maybe a reason for the lack of a
    clear interpretation of the solutions.

\subsection{General features}       

    The solution can be written jointly for $\eps = \pm 1$ if one uses the
    harmonic coordinate $u$ in the metric (\ref{ds_u}), corresponding to the
    coordinate condition $\alpha(u) = \gamma(u) + 2\beta(u)$ \cite{br73}:
\bear                                 \label{fish}
     ds^2 \eql \e^{-2mu}dt^2 - \frac{\e^{2mu}}{s^2(k,u)}
                 \left[\frac{du^2}{s^2(k,u)} + d\Omega^2\right],
\cm
     \phi = Cu,
\ear
    where the integration constants $m$ (the Schwarzschild mass), $C$
    (the scalar charge) and $k$ are related by
\beq                                                       \label{int_0}
            2k^2\sign k = 2m^2 + \eps C^2,
\eeq
    and the function $s(k,u)$ is defined as follows:
\beq                                                       \label{def_s}
    s(k,u) = \vars     {
                    k^{-1}\sinh ku,  \ & k > 0 \\
                                 u,  \ & k = 0 \\
                     k^{-1}\sin ku,  \ & k < 0.  }
\eeq
    The coordinate $u$ is defined in the whole range $u > 0$ for $k\geq 0$
    and in the range $0 < u < \pi/|k|$ for $k < 0$. The value $u=0$
    corresponds to flat spatial infinity, so that at small $u$ the spherical
    radius is $r(u) \approx 1/u$, and the metric is approximately
    Schwarzschild, with $g_{tt} \approx 1 - 2m/r$.

    In case $k > 0$, it is helpful to pass over to the quasiglobal coordinate
    $\rho$ by the transformation
\beq
     \e^{-2ku} = 1 - 2k/\rho \equiv P(\rho),                \label{def_P}
\eeq
    and the solution takes the form
\bear                                                       \label{ds+}
     ds^2 \eql P^{a} dt^2 - P^{-a} d\rho^2
                                 - P^{1 - a} \rho^2 d\Omega^2,
\cm 
     \phi = -\frac{C}{2k} \ln P(\rho),
\ear
    with the constants related by
\beq                                                        \label{int_0+}
    a = m/k, \cm  a^2 = 1 -\eps C^2/(2k^2).
\eeq

    {\bf Fisher's solution} \cite{fisher} corresponds to $\eps = +1$, hence
    according to (\ref{int_0}), it consists of a single branch $k > 0$
    and, in (\ref{ds+}), $|a| < 1$. It is defined in the range $\rho > 2k$,
    while $\rho = 2k$ is a naked central ($r=0$) singularity which is
    attractive for $m >0$ and repulsive for $m < 0$. The Schwarzschild
    solution is restored by making $C=0$, $a = 1$ for $m > 0$ and by $C=0$, $a=-1$
    for $m < 0$.

    The solution for $\eps = -1$ (that is, for a phantom scalar field) may
    be conveniently termed {\bf the anti-Fisher solution}, by analogy with de
    Sitter and anti-de Sitter. Its properties are more diverse and
    interesting. According to three variants of the function (\ref{def_s}),
    this solution splits into three branches with the following main
    properties.

\medskip\noi
{\bf (a)} $k > 0$: the solution again has the form (\ref{ds+}), but now
    $|a| > 1$.  For $m <0$, that is, $a <-1$, we have, just as in the Fisher
    solution, a repulsive central singularity at $\rho = 2k$.

    The situation is, however, drastically different for $a > 1$. Indeed,
    the spherical radius $r$ then has a finite minimum at $\rho = \rho_{\rm
    th} = (a+1)k$, corresponding to a throat of the size
\beq
    r(\rho_{\rm th}) = r_{\rm th} = k (a+1)^{(a+1)/2} (a-1)^{(1-a)/2},
\eeq
    and tends to infinity as $\rho\to 2k$. Moreover, for $a=2,\ 3,\ \ldots$
    the metric exhibits a horizon of order $a$ at $\rho = 2k$ and admits a
    continuations to smaller $\rho$. A peculiarity of such horizons is
    their infinite area. Such horizons have been termed type B \cite{k1,k2}
    horizons, to distinguish them from ordinary, ``type A'' horizons of
    finite area. The whole \asflat\ configurations with type B horizons were
    named cold \bhs\ (CBHs) since all of them have zero Hawking temperature.

    Furthermore, all \Kr\ scalar constituents $K_i$ (see \eqs (\ref{Riem}) in
    the Appendix) behave as $P^{a-2}$ as $\rho\to 2k$ and $P\to 0$.
    An exception is the value $a = 1$, in which case $C=0$, $\phi \equiv
    0$, and the Schwarzschild solution is reproduced. Hence, at $\rho = 2k$ the
    metric has a curvature singularity if $a < 2$ (except for $a=1$), a finite
    curvature if $a=1$ and $a=2$ and zero curvature if $a > 2$.

    For non-integer $a > 2$, the qualitative behavior of the metric as
    $\rho \to 2k$ is the same as near a type B horizon, but a continuation
    beyond it is impossible due to non-analyticity of the function
    $P^a(\rho)$ at $\rho = 2k$. Since geodesics terminate there at a finite
    value of the affine parameter, this is a space-time singularity (a {\it
    singular horizon\/} as it is named in the Appendix) even though the
    curvature invariants tend there to zero.

\medskip\noi
{\bf (b)} $k = 0$: the solution is defined in the range $u \in \R_+$ and is
    rewritten in terms of the quasiglobal coordinate $\rho = 1/u$ as follows:
\bear
       ds^2 = \e^{-2m/\rho}dt^2                                \label{ds_0}
                - \e^{2m/\rho}[d\rho^2 + \rho^2 d\Omega^2],
\cm
       \phi = C/\rho.
\ear
    As before, $\rho = \infty$ is a flat infinity, while at the other
    extreme, $\rho\to 0$, the behavior is different for positive and
    negative mass. Thus, for $m < 0$, $\rho =0$ is a singular center ($r=0$
    and all $K_i$ are infinite). On the contrary, for $m > 0$, $r \to
    \infty$ and all $K_i \to 0$ as $\rho \to 0$. This is again a singular
    horizon: despite the vanishing curvature, the non-analyticity of the
    metric in terms of $\rho$ makes its continuation impossible.

\medskip\noi
{\bf (c)} $k < 0$: the solution describes a \wh\ with two flat
    asymptotics at $u=0$ and $u = \pi/|k|$. The metric has the form
\bear                                                            \label{ds-}
    ds^2 \eql \e^{-2mu} dt^2 - \frac{k^2\e^{2mu}}{\sin^2 (ku)}
        \biggl[ \frac{k^2\,du^2}{\sin^2 (ku)} + d\Omega^2 \biggr]
\nn
       \eql \e^{-2mu} dt^2 - \e^{2mu} [d\rho^2 - (k^2 + \rho^2) d\Omega^2],
\ear
    where $u$ is expressed in terms of the quasiglobal coordinate $\rho$,
    defined on the whole real axis $\R$, by $|k| u =  \cot^{-1} (\rho/|k|)$.
    If $m > 0$, the \wh\ is attractive for ambient test matter at the first
    asymptotic ($\rho\to \infty$) and repulsive at the second one ($\rho \to
    - \infty$), and vice versa in case $m < 0$.  For $m = 0$ one obtains the
    simplest possible \wh\ solution, called the Ellis \wh, although Ellis
    \cite{h_ell} actually discussed these \wh\ solutions with any $m$.

    The \wh\ throat occurs at $\rho = m$ and has the size
\beq
    r_{\rm th} = (m^2 + k^2)^{1/2}
            \exp \left(\frac{m}{k} \cot^{-1}\frac{m}{k}\right).
\eeq

\subsection{Cold \bhs\ in the anti-Fisher solution}

    Among different branches of the anti-Fisher solution, of greatest
    interest for us is the case of CBHs. Let us briefly discuss their
    structure and properties.

    For odd $a$, the principal geometric and causal properties, including the
    Carter-Penrose diagram, coincide with those of the Schwarzschild metric.
    Thus, at $\rho < 2k$, $\rho$ is a temporal coordinate, $t$ spatial, the
    space-time is homogeneous and anisotropic, corresponding to the
    Kantowski-Sachs type of anisotropic cosmologies. The singularity at
    $\rho=0$ ($r=0$) is spacelike (cosmological) and is reached by all
    timelike geodesics in a finite time interval after crossing the horizon.

    For even $a$, the Penrose diagram is the same as that of the extreme
    Reissner-Nordstr\"om space-time; however, the physical meaning of the
    regions where $\rho < 2k$ is quite different. Since $g_{22}$ and $g_{33}$
    change their sign at the horizon, the metric at $\rho < 2k$ has the
    signature $(-\ +\ +\ +)$ instead of $(+\ -\ -\ -)$ at large $\rho$. The
    Lorentzian nature of space-time is still preserved, and one can verify
    that all geodesics are continued smoothly from one region to the other
    (the geodesic equations depend only on the Christoffel symbols and are
    invariant under the anti-isometry $g_{\mu\nu} \to -g_{\mu\nu}$). The
    time coordinate in that region is $\rho$ since $g_{\rho\rho} < 0$ while
    the other diagonal components of $g\mn$ are positive. Thus, just as for
    odd $a$, we have there a Kantowski-Sachs type cosmology with a spacelike
    singularity at $\rho=0$ ($r=0$). The direction of the arrow of time can
    be arbitrary there since timelike geodesics that penetrate from the
    static region become there spacelike (one cannot say for them where is
    the past and where is the future), and can even avoid the singularity.

    The properties of the scalar field are not less exotic. According to
    (\ref{ds+}), $\phi \to \infty$ as $\rho\to 2k$; this, however, does not
    contradict the regularity of the surface $\rho=2k$ for $a\geq 2$ since
    the energy density
\beq
       T^0_0 = -\Half A\phi'{}^2 = -\frac{C^2}{2}\,
                \frac{(\rho-2k)^{a-2}}{\rho^{a+2}},
\eeq
    as well as the other components of $T\mN$, are finite there (recall that
    for $a < 2$ the curvature invariants also diverge, together with
    $T\mN$). Thus the infinite value of $\phi$ does not prevent the
    continuation of the space-time manifold to smaller $\rho$, where the
    solution is valid with $\phi = -(C \ln |P|)/(2k)$. On the other hand,
    the total scalar field energy, calculated as the conserved quantity
    corresponding to the timelike Killing vector, turns out to be infinite
    in the static region independently of $a$:
\beq                                                          \label{E_AF}
    {\cal E} = \int T^0_0\,\sqrt{g} d^3 x
        = -2\pi C^2 \int \frac{d\rho}{\rho(\rho-2k)},
\eeq
    and the integral logarithmically diverges at $\rho=2k$. The divergence
    is related to the infinite spatial volume: the integral $\int \sqrt{^3
    g}d^3 x$ diverges near $\rho = 2k$ even stronger than (\ref{E_AF}).

\section{Comparison with the Brans-Dicke theory. Conformal continuations}

\subsection {Jordan picture in the \BD\ theory}

    The (anti-)Fisher solution, being a solution of general relativity with
    a massless, minimally coupled scalar field, is simultaneously a solution
    of an arbitrary STT in its Einstein picture. Let us discuss the
    corresponding Jordan picture, for certainty, in the context of the
    simplest and most well-known STT, namely, the \BD\ theory. The latter
    corresponds to the choice
\beq
      f(\varphi) = \varphi, \cm               h(\varphi) = \omega/\varphi
\eeq
    in (\ref{act-J}), $\omega$ being the \BD\ coupling constant; we also
    take the massless version of the theory, $U(\varphi) \equiv 0$, to
    deal with counterparts of the (anti-)Fisher solution.

    Since we are only interested in CBHs, let us restrict ourselves to
    solutions with $k > 0$, given by \eq (\ref{ds+}) with (\ref{def_P}).
Then, the Jordan-frame solution of the Brans-Dicke theory
    may be written in the form
\bear                                   \label{metricJ}
    ds^2_J \eql P^{- \xi}\,ds^2_E
    = P^{a - \xi}dt^2 - P^{-a- \xi} d\rho^2 - P^{1 -\xi -a}\rho^2 d\Omega^2,
\\                                                  \label{phiBD}
      \varphi \eql \exp \big[\phi/\sqrt{|\omega+3/2|}\big]
            = P^{\xi},
\ear
    where the parameter $\xi$ is related to $a$ and $\omega$ by
\beq                                                \label{cond}
        (3 + 2\omega)\xi^2 = 1 - a^2.
\eeq

    Conditions for finding black holes in this solution have been discussed
    in Refs.\,\cite{k1,k2}. Let us briefly recall them.

    As in (\ref{ds+}), a horizon in the metric (\ref{metricJ}) can occur at
    $\rho = 2k$ if $a > 0$. However, it has been shown \cite{k1,k2} that
    CBH solutions exist only when the parameters $a$ and $\xi$ obey
    the following ``quantization'' conditions:
\beq                                                         \label{m,n}
     a = \frac{m + 1}{m - n} , \cm\ \xi = \frac{m - n - 1}{m - n},
\eeq
    where $m$ and $n$ are positive integers satisfying the inequalities
\beq
        m - 2 \geq n \geq 0 .                                \label{ineq}
\eeq
    The coupling constant $\omega$ should also belong to a discrete set
    of values,
\beq
       2\omega + 3 = - \frac{2m (n+1) -n^2 +1}{(m-n-1)^2} < 0.
\eeq
    Since, for the \BD\ theory, $l(\varphi) = \omega + 3/2$ and $\eps =
    \sign l$ (see (\ref{g-wag})), we find that, just as in the Einstein
    picture, CBHs can only exist with a phantom scalar. The same is true
    for similar configurations with a nonzero electric charge \cite{robson},
    despite a greater number of classes of solutions.

    However, the CBH existence conditions are different in the Einstein and
    Jordan pictures, and the global structures of the complete space-times,
    continued beyond the horizons, are also different \cite{k1,k2}.
    In particular, as follows from (\ref{Riem}), all $K_i$ turn to infinity
    as $\rho\to 0$ in the solution (\ref{ds+}). In other words, there is
    always a curvature singularity in the internal region of Einstein-frame
    CBHs (or, which is the same, CBHs with a minimally coupled massless
    phantom scalar field in general relativity). Meanwhile, many of
    Brans-Dicke Jordan-frame CBHs are nonsingular, and some of them have
    another flat asymptotic region beyond the horizon \cite{k1,k2}.

\subsection {Conformal continuations--III}

    By (\ref{m,n}), the Jordan-frame CBHs form a discrete family with
    two integer parameters $m$ and $n$ subject to (\ref{ineq}), while the
    family of Einstein-frame CBHs depends on the single integer parameter
    $a\geq 2$. The conformal mapping (\ref{g-wag}) that connects the two
    frames in some cases converts \bhs\ into \bhs, namely, when $m+1$ is a
    multiple of $a$; according to (\ref{m,n}), the parameter $n$ is then
    expressed as $n = m - (m+1)/a$.

    In general, however, the conformal mapping (\ref{g-wag}) converts CBHs
    in \ME\ into configurations with a singular horizon or a curvature
    singularity in \MJ\ and vice versa. Let us give some examples:

\begin{enumerate} \itemsep 0pt
\item
    In case $n = 0$, $m = 2,3, \ldots$, from (\ref{m,n}) we obtain $a =
    (m+1)/m$, in which case the metric (\ref{ds+}) in \ME\ has a
        curvature singularity at $\rho = 2k$.
\item
    Given $m=4$, $n=2$, we have $a = 5/2$, a singular horizon at
        $\rho = 2k$ in \ME.
\item
    Given $a = 2$, i.e., a CBH in \ME, and $m = 2,4,6,\ldots$, we obtain
    half-integer $n$, hence a singular horizon at $\rho = 2k$ in \MJ.
\end{enumerate}

    In all these cases and similar ones, the mapping (\ref{g-wag})
    establishes a one-to-one correspondence between points of \ME\ and \MJ\
    only in the region $\rho > 2k$, which coincides with the whole manifold
    \ME\ but only a portion of \MJ\ in examples 1 and 2, and vice versa in
    example 3. By definition \cite{vac4}, we are thus dealing with conformal
    continuations (CCs).

    A conformal mapping $\cF(\Omega): \M_1 \mapsto \M_2$ between two
    (pseudo-)Riemannian manifolds $\M_1$ and $\M_2$, parametrized by the
    same coordinates $x^\mu$, is a point-to-point mapping such that the
    respective metrics are related by
            $g^{(2)}\mn = \Omega^2(x^\mu)g^{(1)}\mn$,
    where the function $\Omega^2 (x^\mu)$ (the conformal factor) is assumed
    to be smooth in a certain range of the arithmetic space of the
    coordinates $\{x^\mu\}$. Thus, in general, $\cF(\Omega)$ connects only
    some regions of $\M_1$ and $\M_2$ rather than the whole manifolds, and
    which particular regions, depends on the analytic properties of the
    metrics and the conformal factor $\Omega^2$.

    Among different opportunities, a conformal continuation (CC) from $\M_1$
    to $\M_2$ \cite{vac4} is distinguished by the following circumstance:
    it maps a singular surface in $\M_1$ (so that $\M_1$ terminates there)
    to a regular surface $\Str \in \M_2$, so that $\M_2$ continues beyond
    \Str, to a region where the mapping $\cF(\Omega)$ is not defined.

    In normal STT, with $\eps = +1$, the existence of CCs from \ME\ to \MJ\
    in \ssph\ solutions was found to be a generic phenomenon if the function
    $f(\varphi)$ in (\ref{act-J}) has a simple zero \cite{vac4}. It was also
    concluded that the continued manifolds have generically the structure of
    \whs. Explicit examples of CCs are known in the case of nonminimally
    coupled massless scalar fields in general relativity, treated as STT
    (\ref{act-J}) with $f(\varphi) = 1 - \xi\varphi^2$ ($\xi=\const >0$),
    $h(\varphi) = 1$, $U(\varphi) \equiv 0$ \cite{br73,vac4,bar-vis}.

    Ref.\,\cite{vac4} classified CCs by the nature of the transition surfaces
    \Str:
\begin{description}    \itemsep 0pt
\item[CC-I]
    --- \Str\ is an ordinary regular sphere in $\M_2$,
\item[CC-II]
    --- \Str\ is a Killing horizon of finite area in $\M_2$.
\end{description}
    In our case, we have a third type of conformal continuation:
\begin{description}    \itemsep 0pt
\item[CC-III]
    --- \Str\ is a Killing horizon of infinite area (type B) in $\M_2$.
\end{description}

    One could also classify CCs by the types of singularities in $\M_1$ which
    are removed by the appropriate conformal mapping. Thus, in all cases
    considered in \cite{vac4}, the preimage of $\Str \in \M_2 = \MJ$ in
    the manifold $\M_1 = \ME$ was an attracting centre, being a curvature
    singularity like the one in Fisher's solution. Unlike that, in example
    1, the surface $\rho = 2k$ is an attracting curvature singularity of
    infinite radius $r$ while in example 2 it is a singular horizon, i.e., a
    sphere of infinite radius and zero curvature, where the analyticity of
    the metric is lost. The same is true in example 3, but there a singular
    horizon occurs in \MJ\ and a regular type B horizon in \ME.

    To summarize, in the present study we have found CCs of a new type,
    which, unlike those described in Ref.\,\cite{vac4}, (i) exist in
    anomalous (phantom) STT only, (ii) lead to type B (infinite-area)
    horizons as transition surfaces \Str, (iii) have other types of
    singularities as preimages of \Str, and, finally, (iv) can occur not
    only from \ME\ to \MJ, as in examples 1 and 2, but also from \MJ\ to
    \ME, as in example 3. The latter means that a singularity in the Jordan
    picture corresponds to a regular surface in Einstein's.

    In the case of odd $m$ in example 3 and other similar cases, there are
    CBHs in both pictures, the mapping (\ref{g-wag}) transfers a horizon to
    a horizon, but the global structures are different in different
    pictures, and there is a complicated system of one-to-one
    correspondences between different regions of \ME\ and \MJ, depending on
    the particular values of $a$, $m$ and $n$. This issue may be a subject
    of a separate study, which is beyond our scope here.

\subsection{On thermodynamics of scalar black holes}

    The Hawking temperature of a \bh\ horizon is $\TH = (2\pi k_B)^{-1}
    \kappa$, where $k_B$ is Boltzmann's constant while the surface gravity
    $\kappa$ of the horizon is given by the expression \cite{wald}
\beq                                        \label{kap}
    \kappa = \frac{1}{2}\frac{g'_{00}}
                      {\sqrt{|g_{00}g_{11}|}}\Bigg|_{u = u_h}
       = \Half A'(\rho_h),
\eeq
    where $u=u_h$ is the value of an arbitrary radial coordinate $u$ at the
    horizon, and after the second equality sign we give the corresponding
    expression in terms of the quasiglobal coordinate $\rho$ (see the
    Appendix), $\rho_h$ being its value at the horizon.

    The problem of conformal invariance of the Hawking temperature has been
    addressed in Ref.\,\cite{jacobson}. In this work, it has been stated
    that $\TH$ is the same for black holes obtained from conformally related
    theories under the conditions of staticity and asymptotic flatness.

    Indeed, after a conformal transformation $g\mn = \Omega^2{\tilde g}\mn$,
    where $\Omega$ is a function of $\rho$, the surface gravity
    $\tilde\kappa$ at the surface $\rho = \rho_h$, defined in the new
    manifold with the metric $\tilde g_{\mu\nu}$, is
\beq                                                     \label{conf-kap}
    \tilde\kappa = \kappa + A(\rho_h) \frac{\Omega'}{\Omega}(\rho_h).
\eeq
    We have $\tilde\kappa = \kappa$ (i.e., invariance of Hawking's
    temperature under conformal mappings) if the second term in the r.h.s.
    of (\ref{conf-kap}) is zero. And it is really the case since
    $A (\rho_h) =0$ if the conformal factor $\Omega$ is regular at the
    horizon, i.e., if the conformal transformation is well defined on it.

    However, for the presently discussed mapping between the
    anti-Fisher and \BD\ CBHs, the question of invariance of $\TH$ is
    either meaningless or trivial. Indeed, generically, as we have seen above,
    this transformation does not map a black hole to a black hole,
    and the invariance issue is meaningless. On the
    other hand, there is a general law saying that horizons of infinite area
    are always perfectly ``cold'', i.e., have zero temperature. Hence, in the
    cases where such \bhs\ are in conformal correspondence, the invariance
    properties of their temperature become trivial,

    Finally, the infinite horizon area of the CBHs may suggest, according to
    the well-known relations of \bh\ thermodynamics \cite{wald}, that they
    should have infinite entropy. However, after a closer investigation,
    it has been argued that such \bhs\ must in fact have zero entropy
    \cite{zaslavskii}, which seems to be in a better agreement with a zero
    temperature state.  This means that, for such objects, the law that
    relates the \bh\ entropy with its horizon area, is violated.

\section {Cold \bhs\ with $V \not\equiv 0$}

\subsection{Near-horizon behaviour of the solutions}

    So far we have been discussing CBHs with massless scalar fields.
    However, CBH solution with self-interacting scalar fields $V(\phi)
    \not\equiv 0$) also exist, though under certain restrictions.

    Indeed, consider \eqs (\ref{ne1})--(\ref{ne4i}), let a horizon of
    infinite area ($r\to \infty$) be located at $\rho = 0$, and let us
    approximate the metric functions $A(\rho)$ and $r(\rho)$ at small $\rho$
    by
\beq
        A \sim \rho^a, \cm  r^2 \sim \rho^{-b},
\eeq
    where $a = 2,3,\ldots$ and $b > 0$. Substitution into (\ref{ne3})
    results in $\eps = -1$ (the field is necessarily phantom) and $\phi'\sim
    \rho^{-1}$, hence $|\phi| \sim -\log \rho \to \infty$ as $\rho \to 0$.

    Furthermore, from (\ref{ne4i}) one finds
\beq
       (a+b) \rho^{a+b-1} \approx
            \const\cdot (\rho_0 \rho^{2b} - \rho^{2b+1}).
\eeq
    It follows that there can be two families of solutions,
\bearr
        {\rm I}:\ \ \ \rho_0 \ne 0, \cm  b = a-1,             \label{V1}
\\ \lal
    {\rm II}:\ \ \rho_0 =0, \cm    b = a-2.              \label{V2}
\ear
    Substitution into (\ref{ne2}) and (\ref{ne1}) shows that, for family I,
    both $V$ and $dV/d\phi$ behave as $o(\rho^{a-2})$ (where $a \geq 2$),
    while for family II both $V$ and $dV/d\phi$ are of the order of
    $\rho^{a-2}$, where $a \geq 3$ (the value $a=2$ is ruled out by $b > 0$,
    see (\ref{V2})). This second family is, however, of little interest from
    the CBH viewpoint: an analysis of signs in \eq (\ref{ne4i}) shows that
    such a horizon is only accessible from a T region, where $A(\rho) < 0$,
    and cannot be a \bh\ horizon.

    In both families, $|\phi| \to \infty$ at the horizon, and, since
    at the same time $dV/d\phi \sim V \to 0$, the potential behaves at large
    $|\phi|$ as $V \sim \e^{-c|\phi|}$, $c = \const > 0$. We conclude, in
    particular, that only with such potentials CBH solutions may exist.

    One can verify that in all CBH solutions (they belong to family I) all
    functions behave near the horizon as in the anti-Fisher solution
    (\ref{ds+}) with $a = 2,3,\ldots$ (for comparison, in (\ref{ds+}) one
    should move the origin of the $\rho$ coordinate to the horizon, i.e.,
    replace $\rho \mapsto \rho + 2k$). In fact, the anti-Fisher solution is
    a special case belonging to family I. The potential energy density $V$
    at small $\rho$ is much smaller than the kinetic energy density $-A
    \phi'^2/2$, so that the system behaviour near the horizon is dominated
    by the kinetic term.

    (Family II, on the contrary, can only exist in systems with nonzero
    $V(\phi)$, and in this case the potential and kinetic energy densities
    are of the same order near the horizon.)

    The anti-Fisher asymptotic behavior at small $\rho$ indicates that
    these solutions may be converted to the Jordan frame of any STT in the
    same manner as the anti-Fisher solution. Moreover, the small-$\rho$
    behavior of such Jordan-frame solutions in a given STT will be, in the
    main order of magnitude, the same as in the massless case, $V\equiv 0$.
    Therefore, one can assert that their thermodynamic properties and the
    nature of conformal continuations should also be the same as in the
    massless case.

\subsection {Example}

    To obtain an example of an asymptotically flat CBH solution with
    $V \not\equiv 0$, let us use the inverse problem method (see, e.g.,
    \cite{vac2,pha1}) and suppose
\beq \label{hypothesis}
    r(\rho) = \frac{\rho^2}{\sqrt{\rho^2 - b^2}}, \cm b = \const.
\eeq
    Substituting it into \eq (\ref{ne4i}) and imposing $\rho_0 = 35\,b/16$,
    we obtain
\beq
    A(\rho) = \frac{(\rho - b)^2(24\rho^2 + 37\rho b +
            15b^2)}{24\rho^3(\rho + b)} \ .
\eeq
    It is easy to verify that this expression represents an asymptotically
    flat CBH, with $A = 0$ and $A' = 0$ at the horizon, $\rho = b$.
    The potential is given, from (\ref{ne2}), by
\beq
    V =\frac{b^3 (\rho - b)(7\rho + 4b)}{12\rho^4 (\rho + b)^3}\ .
\eeq
    The potential is zero at the horizon and at infinity. Using \eq
    (\ref{ne3}), we find an explicit expression for the scalar field:
\beq
     \phi = \sqrt{\frac 32}
     \log \frac{x^2 - 1}{5 + x^2 + 2\sqrt{3(x^2 + 2)}} +
      2 \log \frac{\sqrt{2} + \sqrt{x^2 +2}}{x}, \cm  x :=\frac{\rho}{b}\ .
\eeq
    The scalar field tends to a constant at infinity and diverges
    logarithmically at the horizon. However, as in the massless case, the
    scalar field energy density is finite at the horizon. The behavior of
    the potential in terms of the scalar field $\phi$ can only be obtained
    implicitly due to a complicated relation between $\phi$ and the radial
    coordinate $\rho$.

    This solution is an explicit example of a CBH with a self-interacting
    scalar field. The behavior of all functions confirms our general
    consideration in the first part of this section.

\section{Conclusions}

   Scalar-tensor theories, which are in general characterized by a
   non-minimal coupling between gravity and the scalar field, predict the
   existence of exotic black holes, which have an infinite horizon area and
   zero Hawking temperature. A well-known conformal mapping transforms
   any scalar-tensor theory from a large class (the Bergmann-Wagoner
   class) into general relativity minimally coupled to a massless scalar
   field. It had been thought for a long time that no black hole solution
   exists in this Einstein-scalar field system, at least for a massless
   scalar field in vacuum. We have shown here that this is not true, and we
   exhibit a new class of black hole solutions. However, for their existence
   the sign of the kinetic term of the scalar field must be reversed,
   leading to a negative-energy field. As in the scalar-tensor case, the
   ``scalar'' black holes have infinite horizon areas and zero temperature.
   However, the conditions in the parameter space for the existence of such
   black holes are different in Jordan's (non-minimal coupling) and
   Einstein's (minimal coupling) conformal frames, which leads to a new type
   of conformal continuations in the Einstein-frame and Jordan-frame
   manifolds.

   The Einstein frame is common to the whole class (\ref{act-J}) of
   scalar-tensor theories, whereas Jordan frames change from theory to
   theory together with the nonminimal coupling functions. This means that
   the discrete ``quantization'' conditions for the solution parameters,
   providing the existence of cold \bhs, will be different in similar
   solutions of different theories.

   The absence of continuations through certain surfaces of finite (or even
   zero) curvature is a peculiar property of many scalar-tensor solutions,
   indicating a special type of space-time singularities related to
   violation of analyticity, which actually means the divergence of some
   invariants of the metric tensor with derivatives of orders higher
   than two. Physical properties of such singularities and their possible
   regularization by taking into account more general solutions or quantum
   corrections may be of considerable interest.

   The Hawking temperature discussed here is expressed in terms of the
   surface gravity $\kappa$. In a more rigorous treatement, quantum fields
   around such black holes must be considered. This is a delicate point,
   since all black holes studied in this work have zero temperature, which
   is, in principle, a violation of the third law of thermodynamics. For
   cold black holes in the Jordan frame there are anomalies in the
   definition of quantum fields, connected with normalization of quantum
   modes \cite{glauber}. However, no complete study in this sense has been
   performed so far, mainly due to technical difficulties. It would be of
   interest to consider this problem in the context of the ``scalar'' black
   holes presented in this work.

\section*{Appendix}
\def\theequation{A.\arabic{equation}}
\sequ{0}

    Cold \bhs\ (CBHs) actually extend the notion of \bhs\ to infinite
    horizon areas. So let us specify what we understand by a ``black hole
    solution''. For our comparatively simple case of \ssph\ space-times,
    leaving aside more general and more rigorous definitions of horizons
    and \bhs\ (see, e.g., \cite{wald}), we can rely on the following working
    definition. A black hole is a space-time containing (i) a static region
    which may be regarded external (e.g., contains a flat asymptotic), (ii)
    another region invisible for an observer at rest residing in the static
    region, and (iii) a Killing horizon of nonzero area that separates the
    two regions and admits an analytical extension of the metric from one
    region to another. This definition certainly implies that the horizon is
    regular, since otherwise it would be a singularity, belonging to the
    boundary of the space-time manifold, across which there cannot be a
    meaningful continuation.

    We are dealing with the general metric (\ref{ds_u}),
\beq                                                            \label{A1}
     ds^2 = \e^{2\gamma(u)}dt^2 - \e^{2\alpha(u)}du^2 -
                \e^{2\beta(u)} d\Omega^2,
\eeq
    or in terms of the ``quasiglobal'' coordinate $\rho$ under the condition
    $\alpha + \gamma=0$, with the notations $\e^{2\gamma}=\e^{2\alpha} =
    A(\rho)$ and $\e^\beta = r(\rho)$,
\beq                                                            \label{A2}
    ds^2 = A(\rho) dt^2 - \frac{d\rho^2}{A(\rho)} - r^2(\rho)d\Omega^2.
\eeq

    A \bh\ horizon may be represented by a sphere $u = u_h$, or $\rho =
    \rho_h$, at which $g_{00} = \e^{2\gamma} = A = 0 $ and at which all
    algebraic curvature invariants are finite. To check the latter, it is
    sufficient to consider the behaviour of the Kretschmann invariant, given
    by
\beq                                                        \label{Kre}
    K = R^{\mu\nu\lambda\gamma}R_{\mu\nu\lambda\gamma}
                    = 4K_1^2 + 8K_2^2 + 8K_3^2 + 4K_4^2,
\eeq
    where
\def\Dot{\dot{\mathstrut}}
\bear                                                       \label{Riem}
    K_1 \eql {R^{01}}_{01} = - \e^{-\alpha - \gamma}
            \Bigr(\dot\gamma\, \e^{\gamma - \alpha}\Bigl)\Dot
         = -\Half A'';
\nn
    K_2 \eql {R^{02}}_{02} = {R^{03}}_{03} =
            - \e^{-2\alpha}\,\dot\beta\, \dot\gamma
         = - \Half \frac{A'r'}{r};
\nn
    K_3 \eql {R^{12}}_{12} = {R^{13}}_{13} = - \e^{-\alpha - \beta}
            \Bigr(\dot\beta\, \e^{\beta-\alpha}\Bigl)\Dot
         = A\frac{r''}{r} + \Half \frac{A'r'}{r};
\nn
    K_4 \eql {R^{23}}_{23} = \e^{-2\beta} - \e^{-2\alpha}{\dot\beta}{}^2
         = \frac{1}{r^2}\, (Ar'{}^2 - 1),
\ear
    where dots denote $d/du$ and primes $d/d\rho$.

    A Killing horizon (simply a horizon for short) $\rho=\rho_h$ admits a
    continuation to other space-time regions if and only if the function
    $A(\rho)$ behaves near it as $(\rho-\rho_h)^a$, $a \in \N$, and $a$ is
    then called the order of the horizon. This restriction is related to a
    distinguished role of the $\rho$ coordinate: near $\rho=\rho_h$ it
    varies (up to a positive constant factor) precisely as the manifestly
    well-behaved Kruskal-like coordinates used for an analytic continuation
    of the metric \cite{k1,k2}. Hence, using this coordinate (which was
    therefore termed {\it quasiglobal\/} \cite{vac4}), one can ``cross the
    horizons'' preserving the formally static expression for the metric.
    It then also follows that $\rho_h$ is always finite.

    In cases when $A(\rho) \sim (\rho-\rho_h)^a$ and $a$ is a fractional
    number, the space-time cannot be continued due to non-analyticity of the
    metric in terms of well-behaved coordinates. The geodesics also cannot
    be continued beyond the corresponding values of their canonical
    parameters. The sphere $\rho = \rho_h$ is thus a singularity, even if
    all curvature invariants are there finite. Such spheres may be referred
    to as {\it singular horizons\/}, to distinguish them from both regular
    horizons (or, simply, horizons) and curvature singularities.

    In the above \bh\ definition, we have omitted the usual requirement that
    the horizon radius $r(\rho_h)$ and area $4\pi r^2(\rho_h)$ should be
    finite. Admitting $r(\rho_h) = \infty$, one can obtain quite a general
    result:

\medskip\noi
    {\it Any horizon of infinite area has zero surface gravity $\kappa$
    (and hence zero Hawking temperature $\kappa/(2\pi k_{B})$).}

\medskip
    Let us prove it for arbitrary \ssph\ space-times. For the metric
    (\ref{A1}) or (\ref{A2}), the surface gravity (\ref{kap}) is expressed
    as \cite{wald}
\beq
    \kappa = \e^{\gamma-\alpha} |\dot\gamma| = \half A'(\rho),
\eeq
    Hence, a horizon with finite surface gravity corresponds to a simple
    zero of $A$, with $A' \ne 0$, at some finite value of $\rho$. On the
    other hand, the regularity conditions require that all $K_i$ (\ref{Riem})
    should be finite at the horizon. In particular, in the same coordinates,
    $K_2 = -\half A'r'/r$, hence, with $A' \ne 0$, $|K_2| < \infty$ is only
    possible in case $|r'/r| < \infty$, which in turn means that
    $\beta = \log r$ is finite at finite $\rho$. Thus a horizon with finite
    temperature can only occur at a sphere of finite radius $r = \e^\beta$.
    Hence, a horizon with an infinite area can only have zero temperature,
    justifying the term ``cold black hole''.

\subsection* {Acknowledgments}

    KB thanks the colleagues from DF-UFES for hospitality. The work was
    supported by CNPq (Brazil). J.C.F and N.P-N thank also CAPES/COFECUB
    (Brazil-France scientific cooperation) for partial financial support and
    IAP (France) for hospitality during part of elaboration of this
    work.

\end{document}